# Wireless Transmission of Video for Biomechanical Analysis

T. Mirzoev, Ph.D.

*The Department of Information Technology*
*College of Information Technology, Georgia Southern University, Statesboro, GA, 30640*

## ABSTRACT

When there is a possibility to wirelessly stream video over a network, a sophisticated computer analysis of the transmitted video is possible. Such process is used in biomechanics when it is important to analyze athletes' performance via streaming digital uncompressed video to a computer and then analyzing it using specific software such as Arial Performance Analysis Systems or Dartfish. This manuscript presents some approaches and challenges in streaming video as well as some applications of Information Technology in biomechanics. An example of how scientists from Indiana State University approached the wireless transmission of video is also introduced.

## THE NEED FOR WIRELESS

Motion analysis in biomechanics provides important tools for scientists to evaluate and analyze movements that create opportunities for improvements in sports performance. High quality video streaming is already widely available with conventionally wired camcorder to a computer. However, when DV-quality video is being transmitted to a laptop computer, the mobile computer's hardware frequently becomes a limitation in transmission of video. If the video stream could be directed wirelessly to a remote workstation with a sufficient hard drive space, then motion video analysis is possible to be implemented either real time or when required. The wireless transmission of video technology is not new but there are several complications that exist and interfere with a successful video transmission. Commonly, biomechanics specialists utilize multiple video cameras that act as video sources for transmission of video. When wireless transmission of video from multiple video sources is desired, there are several considerations to make.

## MOTION ANALYSIS IN BIOMECHANICS

According to Professor Bruce Martin, the suggestion to use cinematography to analyze motion "may have been suggested by the French astronomer Janssen; but it was first used scientifically by Etienne Marey, who first correlated ground reaction forces with movement and pioneered modern motion analysis" (Martin B., 1999). Motion analysis came a long way from cinematography to multi-camcorder setups with HD resolution. According to Dr. Gideon Ariel, the creator of the APAS "the study of the motion of living things is known as 'Biomechanics' and it has evolved from a fusion of the classic disciplines of anatomy, physiology, physics, and engineering" (Ariel G., 2001). *Figure 1* presents the conventional way of motion analyses in biomechanics. Some techniques used to capture and transmit video include wireless LAN (WLAN) connection from a laptop computer to a remote workstation.

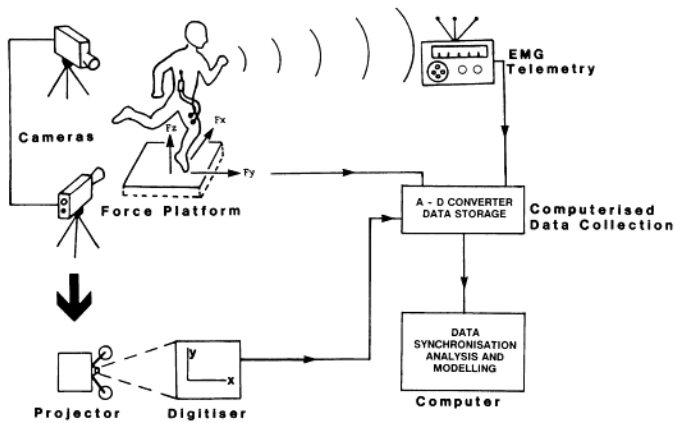

*Figure 1.* A general data acquisition system in biomechanics

Source: "Biomechanical Analysis," Elliott B., Marshall R. (n.d.). Blackwell Science Asia from Bloomfield, Tricker & Fitch, Science and Medicine in Sport. Retrieved from http://www.sportscience.org.nz/publications/guidelines/Section2/2.06_Biomechanical_Analysis.pdf on November 23, 2005.

With rapid development and increased affordability of IT, video analysis in biomechanics is taking a big step towards increased quality of video transmission, precision of video analysis as well as towards increased mobility. When several video cameras are utilized for 3D motion analysis, a typical camera setup could be similar to the one presented on Figure 2.

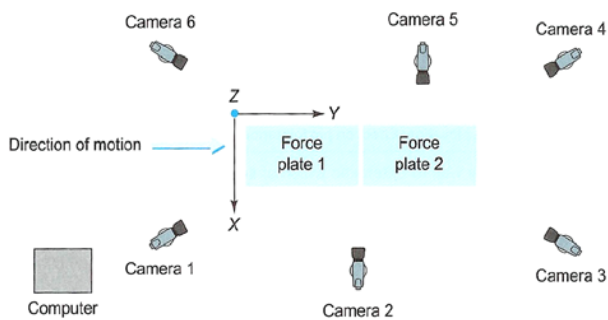

*Figure 2.* Typical camera setup for a 3-D analysis.

Source: Robertson G., Caldwell G., Hamill J., Kamen G., Whittlesey S. (2004). Research Methods in Biomechanics. Human Kinematics, p.147.

A technique with three or more DV-camcorders is used when a 3-D analysis of video is desired. According to Allard, Blanchi and Aissaoui (1995), three-dimensional (3-D) biomechanical analyses "start with data capture by an imaging device". Object points are specified that allow for distinction between different video sources (Allard P., Strokes I., Blanchi J., 1995). Once a video is captured, then computer modeling of human movements is possible. The importance of computer simulation of human movement is stressed by Hamill and Whittlesey. They argue that computer models provide the following advantages (Robertson G., et al. 2004):

1. When dealing with computer models, the constraints associated with human subjects are eliminated: fatigue, strength and coordination, safety and ethics,
2. A computer model allows for experimenting with conditions that cannot be tested on human subjects,
3. Computer models can be adapted to search for optimal solutions.

Besides several advantages that computer models provide there are many limitations that remain, including (Robertson G., et al. 2004):

1. Numerical imperfections exist in the solution process,
2. Difficulty in modeling impacts,
3. Approximation of complex human structures limits computer models,
4. Computer models cannot be compared to enormous adaptability of the human system,
5. Assumptions are a must, since it is not possible to account for all human and environmental factors.

Biomechanical analysis is the evaluation of technique, whether in sports, industry or everyday life. Methods of analysis used in biomechanics vary from those requiring expensive and complex equipment to techniques utilizing little else than an acute eye and an understanding of the mechanics of the movement (Elliott B., Marshall R., n.d.).

## TYPES OF VIDEO STREAMING APPROACHES

According to Mei-Hsuan Lu, Peter Steenkiste, and Tsuhan Chen at Carnegie Mellon University, some of the video streaming problems when applied in WLAN (802.11a) environment could be resolved via Content-Aware Adaptive Retry technique that allows a server to filter out video fields based on a date stamp of each frame (Lu M., Steenkiste P., Chen T., 2005). A sample transmission diagram for

Content-Aware Adaptive Retry is shown by Figure 3.

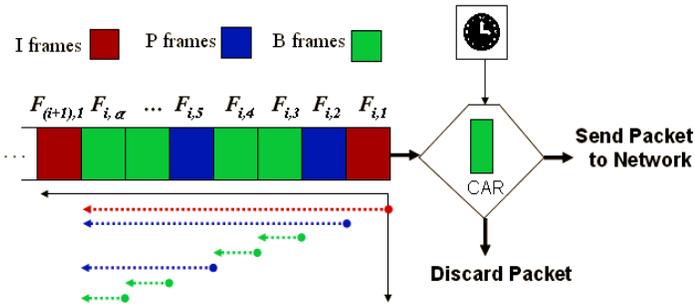

Figure 3. Content-Aware Adaptive Retry architecture.

Source: "Video Streaming Over 802.11 WLAN with Content-Aware Adaptive Retry", Lu M., Steenkiste P., Chen T. (2005). Carnegie Mellon University. Retrieved on September 10, 2005 from http://www.cs.cmu.edu/afs/cs/project/cmcl/archive/2005/icme05.pdf

In this approach, scientists at Carnegie Melon University propose to assign a retransmission deadline to each packet according to its temporal relationship and then make a decision based on the retransmission deadline whether to proceed to network with a packet or disregard it (Lu M., Steenkiste P., Chen T., 2005). Another approach to

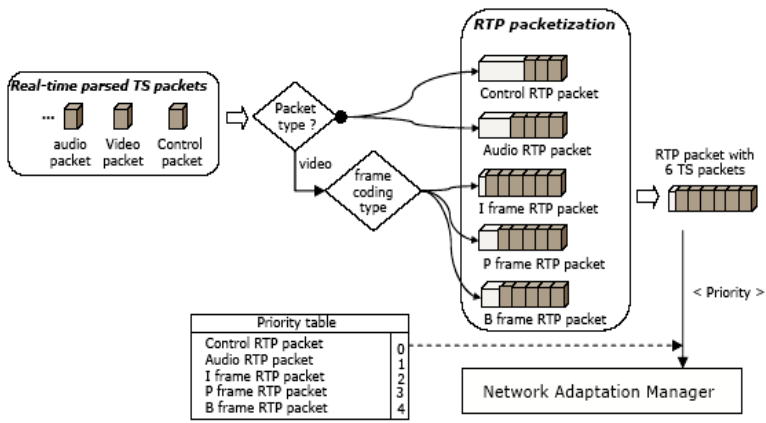

Figure 4. Prioritized real-time transport protocol (RTP) packetization.

Source: From "Network Adaptive High Definition MPEG2 Streaming over IEEE 802.11a WLAN using Frame-based Prioritized Packetization", Lee S., Kim J., Park S. (2005).

video streaming called Frame-based Prioritized Packetization is introduced by Park, Lee and Kim of the Department of Information and Communication Gwangju Institute of Science and Technology (GIST), Korea. The sample of the Frame-based Prioritized Packetization is shown in *Figure 4*. When dealing with wireless transmission of HD video, HD MPEG-2 streaming framework over WLAN is introduced by the scientists from Korea. *Figure 5* presents the HD MPEG-2 streaming framework over WLAN (Lee S., Kim J., Park S., 2005). "To enable the frame-based prioritized packetization, each TS packet is parsed in real-time to find the frame coding type of payload" (Lee S., Kim J., Park S., 2005). In this approach, the scientists from Gwangju Institute of Science and Technology (GIST) propose creating a parsing tool that enables frame-based prioritized packetization over the IEEE 802.11a WLAN (Lee

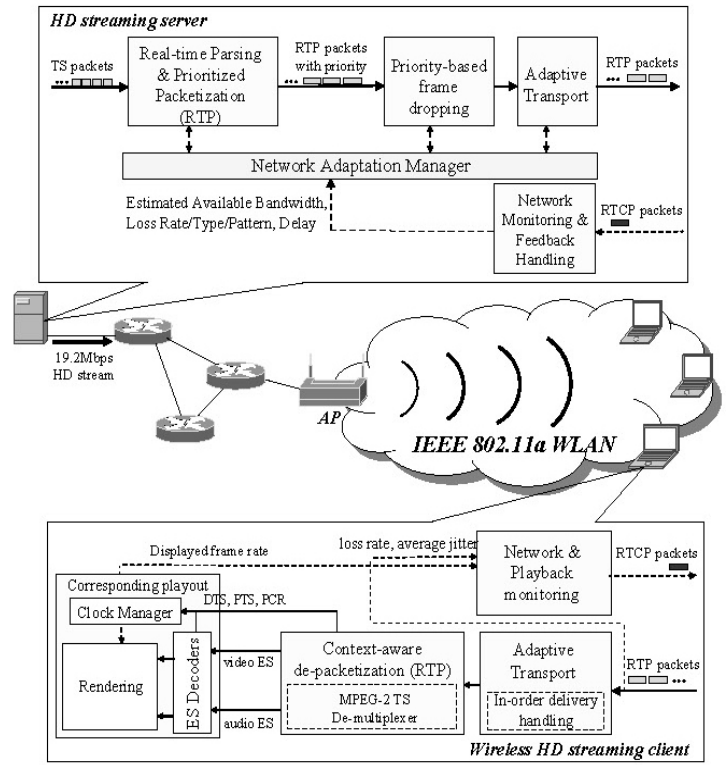

S., Kim J., Park S., 2005).

Figure 5. HD MPEG-2 streaming framework over WLAN

Source: "Network Adaptive High Definition MPEG2 Streaming over IEEE 802.11a WLAN using Frame-based Prioritized Packetization", Lee S., Kim J., Park S. (2005).

## SOFTWARE APPLICATIONS

One of the common software packages used for analysis of motion is called motion analysis software package called Ariel Performance Analysis Software (APAS). This software allows direct connection between several video sources and a PC-based system. The commonly used ports for transmission of video are the USB 2.0 and IEEE 1394a connectors. APAS is a powerful tool that allows for detailed analysis of motion video which then results in suggestions for improvements in the areas where video analysis is desired. According to www.arielnet.com the following are some facts about APAS (Ariel G., 2001):

> Biomechanical quantification are based on Newtonian equations and the APAS analytic technique models the human body as a mechanical system of moving segments upon which muscular, gravitational, inertial, and reaction forces are applied. Although the system has primarily been used for quantification of human activities, it has had some industrial, non-human applications. The computerized hardware/software technique provides a means to objectively quantify the dynamic components of movement and replaces mere observation and supposition.
>
> Ariel Dynamics Inc. invented the first computerized Movement Analysis System, known as the Ariel Performance Analysis System (APAS) in 1968. The system's inventor, Dr. Gideon Ariel, developed the first online electronic digitizing system for reducing each picture in a film sequence, and later from a video, into its kinematic components.
>
> Since 1971, the Ariel Performance Analysis System has assisted medical professionals, sport scientists, and athletes to understand and analyze movement using its advanced video and computer motion measurement technology. It surpasses all other video systems for quantitative accuracy and scientific excellence for the most cost effective choice in major medical and research institutions around the world.

Figure 6 shows an example of how APAS could be used for biomechanical analysis of DV- quality video.

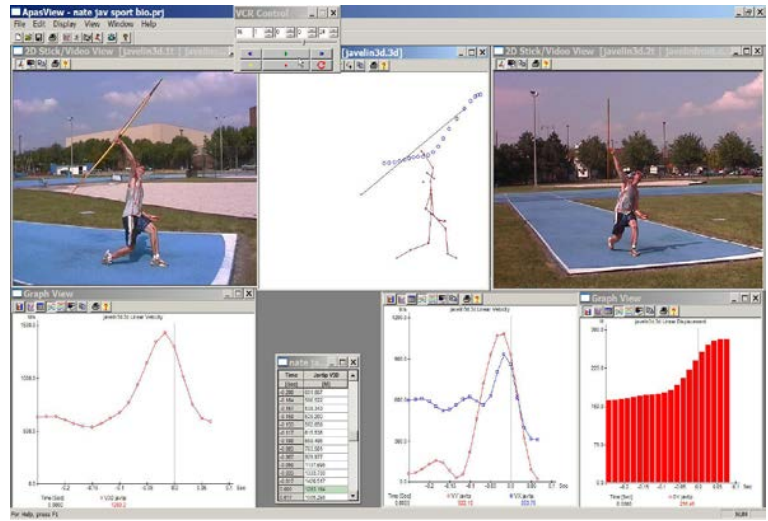

Figure 7. Sample of video based biomechanical analysis with multiple video views.

Source: Indiana State University Biomechanics Laboratory

## IMPLEMENTATION OF WIRELESS VIDEO TRANSMISSION AT INDIANA STATE UNIVERSITY

Dr. Finch and Dr. Mirzoev from Indiana State University (ISU) setup the wireless transmission of video for biomechanical analysis via the existing 802.11a network at ISU. With help from Office of Information Technology Department at ISU, the following was achieved:

    1) Three mini-DV camcorders connected to a laptop computer with installed Ariel Performance Analysis Software provided the transmission of video,

    2) A Gateway 450ROG computer was utilized to receive multiple video streams and cache a video file to its hard drive,

    3) IEEE 802.11a ISU Novell network was used to transmit the file to a remote computer,

    4) A remote Dell Precision Workstation computer was used to receive and store the transmitted video file.

Three video camcorders were connected to a Gateway laptop computer which only cached the transmitted via IEEE 1394 connection video file. The actual writing of the file was performed on a remote Dell Workstation with an allocated partition. A Novell network share set up on the workstation allowed to directly write on Dell's hard drive once a user logged in to the laptop computer via Novell directory that automatically created authenticated user's network shares. Wirelessly, APAS then wrote the cached on the

Gateway laptop video file onto the remote workstation.

The example described here is based on the actual project results that allow students and faculty from Indiana State University analyze motion video in the ISU's biomechanics laboratory.

## CONCLUSION

Besides wired area networks technologies advancement, there is a rapid development of wireless technologies and many new applications continue to enter technology markets. Voice over IP, video over IP, EDGE, GSM, GPRS and satellite networks provide powerful technologies via wireless communications. Streaming high quality video is one of the most demanding technologies since it requires a lot bandwidth and specialized software support. However, despite the demands of video streaming, according to Carroll and Kaven, video streaming is growing rapidly in many diverse areas (Kaven, Carroll, 2000):

> CNN, MSNBC, and TechTV, for instance, all provide video feeds through distributed multimedia servers. Current financial information streams to brokers, and increasingly, to online traders. Education is also moving toward the cutting edge, as canned and live virtual classrooms bring dispersed students together. Companies stream advertising, corporate communications, presentations, slide shows, and sales force training videos over the Internet and their corporate intranets. Instructional videos lower customer-support costs, creating bottom-line business benefits. Streaming content can be found on community sites such as AOL and MindSpring, and on portals like Yahoo! and ZDNet.

Video streaming is becoming common even on mobile phones, which in their turn utilize new technologies at full speed of research and development. The analysis of video for biomechanical analysis is another application that utilizes wireless video streaming technology.

## BIBLIOGRAPHY

Allard P., Strokes I., Blanchi J., (1995). Three Dimensional Analysis of Human Movement. Human Kinetics.

Ariel Dynamics (2001). Description - Ariel Performance Analysis System, Ariel Dynamics Inc. Retrieved on November 25 from 2005http://www.sportsci.com/adi2001/adi/products/apas/system/description.asp#destHeader275.

Carroll S., Kaven O., (2000). Delivering Streaming Video, PC Magazine, 0888-8507, October 3, Vol. 19, Issue 17.

Elliott B., Marshall R. (n.d.) Biomechanical Analysis, Blackwell Science Asia from Bloomfield, Tricker & Fitch, Science and Medicine in Sport. Retrieved from http://www.sportscience.org.nz/publications/guidelines/Section2/2.06_Biomechanical_Analysis.pdf on November 23, 2005.

Lee S., Kim J., Park S. (2005). Network Adaptive High Definition MPEG2 Streaming over IEEE 802.11a WLAN using Frame-based Prioritized Packetization. Department of Information and Communication Gwangju Institute of Science and Technology (GIST), Korea.

Lu M., Steenkiste P., Chen T. (2005). Video Streaming Over 802.11 WLAN with Content-Aware Adaptive Retry. Carnegie Mellon University. Retrieved on September 10, 2005 from http://www.cs.cmu.edu/afs/cs/project/cmcl/archive/2005/icme05.pdf.

Martin B. (1999). A Genealogy of Biomechanics. 23rd Annual Conference of the American Society of Biomechanics University of Pittsburgh, Pittsburgh PA. October 23, 1999.

Robertson G., Caldwell G., Hamill J., Kamen G., Whittlesey S. (2004). Research Methods in Biomechanics. Human Kinematics.

Zeng J., Lee M. (2004). A Comprehensive Performance Study of IEEE 802.15.4. The City University of New York, NY, Department of Electrical Engineering.